\documentclass[twocolumn,pra,multicol,amsmath,amssymb,superscriptaddress]{revtex4-1}
\usepackage{graphicx}
\usepackage{epstopdf}
\usepackage{bm}
\usepackage{subfigure}
\usepackage{sidecap}
\usepackage {times}
\usepackage{ulem}
\usepackage{graphicx}
\usepackage{epstopdf}
\usepackage{amssymb}
\usepackage{amsmath}

\usepackage{bm}
\graphicspath{{Figs/}}

\usepackage{subfigure}
\usepackage{sidecap}

\newcommand\redout{\bgroup\markoverwith{\textcolor{red}{\rule[.5ex]{2pt}{0.4pt}}}\ULon}

\usepackage[colorlinks=true,citecolor=blue]{hyperref}
\hypersetup{colorlinks=true,citecolor=blue,linkcolor=red,urlcolor=blue}

\newcommand{\be}{\begin{equation}}
\newcommand{\ee}{\end{equation}}

\newcommand{\bea}{\begin{eqnarray}}
\newcommand{\eea}{\end{eqnarray}}
\newcommand{\bd}{\begin{displaymath}}
\newcommand{\ed}{\end{displaymath}}
\newcommand{\ba}{\begin{array}}
\newcommand{\ea}{\end{array}}
\newcommand{\bi}{\begin{itemize}}
\newcommand{\ei}{\end{itemize}}
\newcommand{\bc}{\begin{center}}
\newcommand{\ec}{\end{center}}
\newcommand{\bfl}{\begin{flushleft}}
\newcommand{\efl}{\end{flushleft}}
\newcommand{\bfr}{\begin{flushright}}
\newcommand{\efr}{\end{flushright}}

\newcommand{\bl}{\begin{aligned}}
\newcommand{\el}{\end{aligned}}

 \def\bd{{\bf d}}

\def\bra{\langle}
\def\ket{\rangle}

\begin{document}

\title{Dynamics of Quantum Coherence and Quantum Fisher information After a Sudden Quench}

\author{R. Jafari}
\email{rohollah.jafari@gmail.com}
\affiliation{Department of Physics, Institute for Advanced Studies in Basic Sciences (IASBS), Zanjan 45137-66731, Iran}
\affiliation{Department of Physics, University of Gothenburg, SE 412 96 Gothenburg, Sweden}
\affiliation{Beijing Computational Science Research Center, Beijing 100094, China}
\author{Alireza Akbari}
\email{akbari@postech.ac.kr}
\affiliation{Max Planck Institute for the  Chemical Physics of Solids, D-01187 Dresden, Germany}
\affiliation{Max Planck POSTECH Center for Complex Phase Materials, and Department of Physics, POSTECH, Pohang, Gyeongbuk 790-784, Korea}
\affiliation{Department of Physics, Institute for Advanced Studies in Basic Sciences (IASBS), Zanjan 45137-66731, Iran}

\date{\today}

\begin{abstract}
The dynamics of relative entropy and $l_{1}$-norm of coherence, as well as, the Wigner-Yanase-skew
and quantum Fisher information are studied in the one dimensional XY spin chain in the presence of a time-dependent
transverse magnetic field.
We show that independent of the initial state of the system and while the  relative entropy of coherence, $l_{1}$-norm of coherence,
and quantum Fisher information are incapable, surprisingly, the Wigner-Yanase-skew information dynamic can truly spotlight
the equilibrium critical point.
We also observe that when the system is quenched to the critical point, these  quantities show suppressions
and revivals. Moreover, the first suppression (revival) time scales linearly with the
system size and  its scaling ratio is unique for all quenches independent to the initial phase of the system.
This is the promised universality of the first suppression (revival) time.
\end{abstract}
\maketitle
\section{Introduction}
The evaluating quantum coherence (QC) is highly substantial for both quantum foundations and quantum technologies~\cite{Horodecki:2009aa,Modi:2012aa}.
 Quantum coherence itself  represents an essential feature of quantum states and supports all forms of quantum correlations~\cite{Ficek},
however the inevitable interaction of the system with the environment mostly brings incoherency to the input states and evolves a coherence loss~\cite{Chuang}.
Recently, several precise measures have been introduced to quantify the quantum coherence~\cite{Baumgratz,Xi,Hu,Qin}, including
the $l_{1}$-norm quantum coherence  (C$l_{1}$)~\cite{Baumgratz,Balazadeh2020}, the relative entropy of coherence (REC)~\cite{Baumgratz}, the trace norm quantum coherence
(TQC)~\cite{Baumgratz,Rana} and the Wigner-Yanase skew (WYSI) information~\cite{Girolami}.
Among these quantum resource measures,  TQC and C$l_{1}$ 
are defined through a well trace norm, where a closed analytical formula for calculating X-states has been derived invariant under unitary transformations~\cite{Rana, Huang,Huang:2017aa,Lei,Mzaouali}.
Skew information firstly introduced by Wigner and Yanase in 1963~\cite{Wigner910}, and it was originally used to
represent the information content of mixed states. 
In the theory of statistical estimation, the statistical idea govern skew information is the Fisher information~\cite{Luo}, which is not only a key notion of statistical inference~\cite{fisher1925} but also plays an important role in informational treatments of physics~\cite{Helstrom,Frieden,Frieden2}.
\\

Nowadays, quantum Fisher information (QFI), as a witness of multipartite entanglement,
displays much richer aspects of complex structures of topological states~\cite{Zhang}.
 It has been extensively explored in many different fields such as  the calculation of quantum speedup
limit time~\cite{Taddei}; the study of uncertainty relations~\cite{Gibilisco,Karpat};
and the properties of quantum phase transition~\cite{Invernizzi,Sun}.
In particular,  the quantum Fisher information  prepares a bound to characterize the members of a family of probability distributions.
Moreover, when quantum systems are involved, an excellent measurement may be found using tools from quantum estimation theory.
This is especially true for a kind of  problems that the quantity of interest is not directly available.

Quantum Fisher information has introduced
the quantum version of the Cram\'{e}r-Rao inequality~\cite{Helstrom,Holevo,Braunstein,Braunstein:1996aa}
and has imposed the lower bound~\cite{Braunstein}.
Moreover, different features of quantum coherence have been studied, including quantification,
dynamic evolution and operational explanation of quantum coherence~\cite{Hu,Wang,Liu:2016aa,Xueyuan,Yao}.
Some recent works have also examined  the relationship between quantum coherence and quantum phase transition~\cite{Invernizzi,Sun,Karpat},
as well as,
the performance
of the quantum walk version of the Deutsch-Jozsa algorithm and the deterministic quantum computation with one quantum bit (DQC1)  algorithm~\cite{Hillery,Ma,Matera}.
Additionally, it has been shown that multipartite entanglement which witnessed by QFI can capture a quantum phase transition point~\cite{Braunstein2,Giovannetti}.
However, despite several works on  quantum coherence  and QFI, the dynamics of  quantum coherence  and QFI have not yet been studied sufficiently.
Therefore, understanding dynamical
behaviour of  quantum coherence  and QFI would be very useful for the description of the nonequilibrium dynamics and universal behavior of
quantum many-body systems~\cite{Polkovnikov,Pappalardi,Hamma,Jafari1,Jafari2,Jafari3,Mishra2,jafari2019dynamical,Jafari2015,Jafari2010,Jafari2016}.

In this paper, by considering a one-dimensional XY-model with time-dependent (step function) couplings, in an external time-dependent (step function) transverse magnetic field, we study the dynamical behavior of the relative entropy of coherence, $l_{1}$-norm of coherence, and also as measures of quantum coherence, the Wigner-Yanase-skew, and quantum Fisher information.
We  find that, all of these quantities show suppressions and revivals when the system is quenched to the critical point.
We also show that the first suppression (revival) time scales linearly with the system size. This scaling ratio is independent of the size of the quench and the initial preparation phase of the system. 

\section{Time Dependent XY-model}
The Hamiltonian of time-dependent XY-model in a one-dimensional lattice is given by~\cite{Barouch1,Barouch2,Sadiek_2010,Huang2006}
%
\be
\bl
{\cal H}
\!
=
\!\!
-
\!
\sum_{i=1}^{N}
\Big[
J(t)
[
 (1
 \!
 +
 \!
 \gamma)
 S_{i}^{x} S_{i+1}^{x}
 \!
 +
 \!
 (1
  \!
-
 \!
\gamma)
S_{i}^{y} S_{i+1}^{y}
]
\!
+
\!
h(t) S_{i}^{z}
\Big],
\label{eq1}
\el
\ee
%
where $N$ shows the site's number, and $\gamma$ is the anisotropy parameter.
We consider the periodic boundary condition and $S_i^\alpha$ are the  spin half operators at the  $i$th site, which are defined by half of the Pauli matrices as follow
$$S_i^\alpha=\frac{1}{2}\sigma^{\alpha}_{i}; \;\;\; \alpha=\{x,y,z\}.$$
To study the effect of a time-varying coupling parameter, $J(t)$, and magnetic field, $h(t)$, we assume the following
expressions
%
\be
\bl
\label{eq2}
J(t)
&=J_0 + (J_1 - J_0) \Theta(t) ;
\\
h(t)
&=h_0 + (h_1 - h_0) \Theta(t),
\el
\ee
%
with  the Heaviside step function defined by
%
\be
\bl
\label{eq3}
\Theta (t)=\left\{
\begin{array}{lr}
0 & \qquad t\leq 0 \\
1 & \qquad t>0
\end{array}.
\right.
\el
\ee
%
The considered model, Eq.~(\ref{eq1}), can be exactly diagonalized by standard Jordan-Wigner transformation~\cite{Barouch1,Barouch2,Sadiek_2010,Huang2006,Utkarsh}.
Then the Liouville equation of the Fourier transformed Hamiltonian can be solved exactly
and the magnetization and two-point correlation functions can be calculated analytically~\cite{Barouch1,Barouch2,Sadiek_2010,Huang2006,Utkarsh}. 
\\

In the subsequent calculations, we assume that the system is initially at the thermal equilibrium.
In this respect, the reduced two-spin density
matrix $\varrho_{l,m}(t)$  is achieved by
%
%
\be
\bl
\label{eq4}
\varrho_{l,m}(t)=
\left(\begin{array}{cccc}
 \rho_{11} &0 &0 & \rho_{14} \\
 0 & \rho_{22} & \rho_{23} &0 \\
 0 & \rho_{23}^{\ast} & \rho_{33} &0 \\
  \rho_{14}^{\ast} &0 &0 &\rho_{44} \\
\end{array}
\right) ,
\el
\ee
%
where its matrix elements can be written in terms of one- and two-point correlation functions, which are given by
%
%
\be
\label{eq5}
\bl
\rho_{11}
=&
\left\langle M^z_{l}\right\rangle+\left\langle S^{z}_{l} S^{z}_{m} \right\rangle+\frac{1}{4};
\;\;
\rho_{22}
=
\rho_{33}=-\left\langle S^{z}_{l} S^{z}_{m} \right\rangle+\frac{1}{4} ;
\\
\rho_{23}
=&
\left\langle S^{x}_{l} S^{x}_{m} \right\rangle+\left\langle S^{y}_{l} S^{y}_{m} \right\rangle;
\;\;\;\;\;
\rho_{14}
=
\left\langle S^{x}_{l} S^{x}_{m} \right\rangle-\left\langle S^{y}_{l} S^{y}_{m} \right\rangle;
\\
\rho_{44}
=&
- \left\langle M^z_{l}\right\rangle+\left\langle S^{z}_{l} S^{z}_{m} \right\rangle+\frac{1}{4},
\el
\ee
%
with the magnetization in the $z$-direction characterized   as follow
%
\be
\label{mag1}
M^{z}=\frac{1}{N}\sum_{j=1}^{N}
M_{j}^{z}
=\frac{1}{N}\sum_{j=1}^{N}
S_{j}^{z}.
\ee
%
Here, the expectation for the average value is defined by
%
\be
\label{mag2}
\left\langle \cdots \right\rangle=\frac{
{\rm Tr}
[(\cdots)  \rho(t)]}{
{\rm Tr}
[\rho(t)]},
\ee
%
where the exact analytical form of the magnetization, and two point spin-spin correlation functions are
precisely  presented in refs.~\cite{Barouch1,Barouch2,Sadiek_2010,Huang2006,Utkarsh} (see also Appendix \ref{AppA}).
%

\section{Quantum Coherence and Quantum Fisher information}
As mentioned, quantum coherence  is a fundamental physical resource in quantum information tasks~\cite{Streltsov2017}, and
revealing quantum coherence is imperative to accomplish the realization of the quantum correlations.
It is understood as  a key root for  physical resources in quantum computation and quantum information processing, and
a rigorous theory has been proposed to define an excellent notion for measuring it~\cite{Baumgratz}.
In this section we briefly quantify and review the relative entropy of coherence, $l_{1}$-norm of coherence,
Wigner-Yanase-skew information, and the 
quantum Fisher information.
%
%
\subsection{The relative entropy  and $l_{1}$-norm of coherence}
The $l_1$-norm of coherence is defined as a sum of the absolute values of all
off-diagonal elements in the density matrix, $\varrho_{l,m}$, using following expression~\cite{Baumgratz}
%
\be
\bl
\label{eq6}	
{\rm C}l_{1}(\varrho)=\sum_{l\neq m}|\varrho_{l,m}|.
\el
\ee
%
Moreover, the relative entropy of coherence  is defined as
%
\bea
\label{eq7}
{\rm C}_{\rm REC}(\varrho)=S(\varrho_{\rm{diag}})-S(\varrho),
\eea
%
where, $\varrho_{\rm{diag}}$ is the diagonal part of $\varrho_{l,m}$, and the function
\be
S(\varrho_{l,m})=-\rm{Tr}
\Big[
\varrho_{l,m}\log_2\varrho_{l,m}
\Big],
\ee
 is the von Neumann entropy of the density matrix
 $\varrho_{l,m}$.
Calculating the $l_1$-norm for a  transverse field XY-model is straightforward and results
%
\be
\bl
\label{eq8}
{\rm C}l_{1}=4
\;
|\langle S^{x}_{l} S^{x}_{m}\rangle|,
\el
\ee
%
furthermore, using the relative entropy formula, Eq.~(\ref{eq7}), we have
%
\be
\bl
\label{eq9}
{\rm C}_{\rm REC}=\sum_{q=0}^1 (\xi_q\log\xi_q+\eta_q\log\eta_q-\zeta_q\log\zeta_q)-2\varepsilon\log\varepsilon,
\el
\ee
%
with
%
\be
\bl
\label{eq10}
\xi_q
\!
=&
\frac{1}{4}-\langle S^{z}_{l} S^{z}_{m}\rangle +(-1)^q
\Big(
\langle S^{x}_{l} S^{x}_{m}\rangle+\langle S^{y}_{l} S^{y}_{m}\rangle
\Big),
\\
\eta_q
\!
=&
\frac{1}{4}+\langle S^{z}_{l} S^{z}_{m}\rangle
+(-1)^q
\sqrt{
\!
\langle S^{z}_{l}\rangle^2+(\langle S^{x}_{l} S^{x}_{m}\rangle
\!
-
\!
\langle S^{y}_{l} S^{y}_{m}\rangle)^2},
\\
\varepsilon
\!
=&\frac{1}{4}-\langle S^{z}_{l} S^{z}_{m}\rangle.
\el
\ee
%

\subsection{The Wigner-Yanase-Skew Information}{\label{WYSI}}

The definition of the Wigner-Yanase-Skew Information which used as a measure of quantum coherence is given by~\cite{Wigner910,Girolami,Karpat,Gedik}
%
\bea
\bl
I(\varrho,V)=-\frac{1}{2}
 {\rm Tr}
[\sqrt{\varrho},V]^{2},
\el
\eea
%
where the density matrix $\varrho$ depict a mixed quantum state, $V$ is an
observable, and $[\cdots,\cdots]$ represents the commutator.
The quantity $I(\varrho,V)$ can also be interpreted as a measure of the quantum uncertainty of
$V$ in the state $\varrho$ instead of the conventional variance.
A set of the local spin's elements ($S^{\alpha}$)  is an arbitrary and natural choice of observable which constitutes an
local orthonormal basis, as
%
\be
{\rm LQC}_\alpha= I(\varrho_{l,m},S^{\alpha}_{l}\otimes \openone_{m}).
\ee
%
The reduced two-spin density matrix, Eq.~(\ref{eq4}), facilitates the analytical evaluation of the Wigner-Yanase skew information
of the two-spin density matrix. 
Thus, one can  obtain  the eigenvalues and their corresponding normalized eigenvectors of the density
matrix as 
%
\be
\bl
\label{11}
&
p_{1}=\frac{1}{2}(\rho_{11}+\rho_{44}+\sqrt{(\rho_{11}-\rho_{44})^{2}+4|\rho_{14}|^{2}}),\\
&
p_{2}=\frac{1}{2}(\rho_{11}+\rho_{44}-\sqrt{(\rho_{11}-\rho_{44})^{2}+4|\rho_{14}|^{2}}),\\
&
p_{3}=\frac{1}{2}(\rho_{22}+\rho_{33}+\sqrt{(\rho_{22}-\rho_{33})^{2}+4|\rho_{23}|^{2}}),\\
&
p_{4}=\frac{1}{2}(\rho_{22}+\rho_{33}-\sqrt{(\rho_{22}-\rho_{33})^{2}+4|\rho_{23}|^{2}}),
\el
\ee
%
and
%
\be
\bl
\label{eq12}
&
|\phi _{1}\rangle=\frac{1}{N_{1}}\left(
                       \begin{array}{c}
                         \rho_{14} \\
                         0 \\
                         0 \\
                         p_{1}-\rho_{11} \\
                       \end{array}
                     \right);
                     \;\;\;
|\phi _{2}\rangle=\frac{1}{N_{2}}\left(
                       \begin{array}{c}
                         \rho_{14} \\
                         0 \\
                         0 \\
                         p_{2}-\rho_{11} \\
                       \end{array}
                     \right);
                     \\
                     \\
&
|\phi _{3}\rangle=\frac{1}{N_{3}}\left(
                       \begin{array}{c}
                         0 \\
                         \rho_{23} \\
                          p_{3}-\rho_{22} \\
                         0 \\
                       \end{array}
                     \right);
                     \;\;\;
|\phi _{4}\rangle=\frac{1}{N_{4}}\left(
                       \begin{array}{c}
                         0 \\
                         \rho_{23} \\
                          p_{4}-\rho_{22} \\
                         0 \\
                       \end{array}
                     \right),
\el
\ee
%
respectively.
Here $N_{i}$ ($i = 1,2,3,4$) are the normalization factors  defined  by
%
\be
\bl
\label{eq13}
N_{1}
\!
\!
=&
\sqrt{
|\rho_{14}|^{2}
\!
+
\!
(p_{1}-\rho_{11})^{2}};
\;\;
N_{2}
=
\!
\!
\sqrt{|\rho_{14}|^{2}
\!
+
\!
(p_{2}-\rho_{11})^{2}};
\\
N_{3}
\!
\!
=&
\sqrt{|\rho_{23}|^{2}
\!+
\!
(p_{3}
\!
-
\!
\rho_{22})^{2}};
\;\;\;
N_{4}
=
\!
\!
\sqrt{|\rho_{23}|^{2}+(p_{4}-\rho_{22})^{2}}
.
\el
\ee
%
By straightforward calculations, the root of the two-qubit reduced state $\sqrt{\varrho_{l,m}}$
can be obtained by
%
\be
\bl
\label{eq14}
\sqrt{\varrho_{l,m}}=\left(
                       \begin{array}{cccc}
                         \alpha_\varrho & 0 & 0 & \lambda_\varrho \\
                         0 & \beta_\varrho & \nu_\varrho & 0 \\
                         0 & \nu_\varrho^{\ast} & \gamma_\varrho & 0 \\
                         \lambda_\varrho^{\ast} & 0 & 0 & \delta_\varrho \\
                       \end{array}
                     \right),
                     \el
\ee
%
with the following elements
%
\be
\bl
\label{eq15}
\alpha_\varrho
=&
|\rho_{14}|^{2}\Big(\frac{\sqrt{p_{1}}}{N_{1}^{2}}+\frac{\sqrt{p_{2}}}{N_{2}^{2}}\Big),
\\
\beta_\varrho
=&
|\rho_{23}|^{2}\Big(\frac{\sqrt{p_{3}}}{N_{3}^{2}}+\frac{\sqrt{p_{4}}}{N_{4}^{2}}\Big),
\\
\gamma_\varrho
=&
\frac{\sqrt{p_{3}}(p_{3}-\rho_{22})^{2}}{N_{3}^{2}}+\frac{\sqrt{p_{4}}(p_{4}-\rho_{22})^{2}}{N_{4}^{2}},
\\
\delta_\varrho
=&
\frac{\sqrt{p_{1}}(p_{1}-\rho_{11})^{2}}{N_{1}^{2}}+\frac{\sqrt{p_{2}}(p_{2}-\rho_{11})^{2}}{N_{2}^{2}},
\\
\lambda_\varrho
=&
\rho_{14}\Big(\frac{\sqrt{p_{1}}(p_{1}-\rho_{11})}{N_{1}^{2}}+\frac{\sqrt{p_{2}}(p_{2}-\rho_{11})}{N_{2}^{2}}\Big),
\\
\nu_\varrho
=&
\rho_{23}\Big(\frac{\sqrt{p_{3}}(p_{3}-\rho_{11})}{N_{3}^{2}}+\frac{\sqrt{p_{4}}(p_{4}-\rho_{11})}{N_{4}^{2}}\Big).
\el
\ee
%
Along, for the bipartite system in Eq.~(\ref{eq4}), 
the two-spin local
quantum coherence (LQC) components can be written as~\cite{li2013}
%
\be
\bl
\label{eq16}
{\rm LQC}_x
=&1-2(\alpha_\varrho \beta_\varrho +\gamma_\varrho \delta_\varrho )-4{\rm Re}
\Big[
\lambda_\varrho \nu_\varrho
\Big],\\
{\rm LQC}_y
=&1-2(\alpha_\varrho \beta_\varrho +\gamma_\varrho \delta_\varrho)+4
{\rm Re}
\Big[
\lambda_\varrho \nu_\varrho
\Big],\\
{\rm LQC}_z
=&1-
\Big[
\alpha_\varrho^{2}+\beta_\varrho^{2}+\gamma_\varrho^{2}+\delta_\varrho^{2}-2
\Big(
|\lambda_\varrho|^{2}+|\nu_\varrho|^{2}
\Big)
\Big],
\el
\ee
%
which quantify the coherence with respect to the first subsystem locally.

\subsection{ The Quantum Fisher information}
Estimation theory is an important topic in different areas of physics~\cite{Giovannetti,Chin,Holevo,Helstrom,Liu,Jin}.
In general phase estimation perspective, the evolution of a mixed quantum state, given by the density matrix $\varrho$,
under a unitary transformation, can be described as
\be
\varrho_\theta=e^{-i A\theta}\varrho \; e^{i A\theta},
\ee
where $\theta$ is the phase shift and $A$ is an operator.
The estimation accuracy for $\theta$ is bounded by the quantum Cram\'{e}r-Rao inequality~\cite{Helstrom,Holevo}:
%
\bea
\Delta\hat{\theta}\ge\frac{1}{\sqrt{\nu \mathcal{F}(\varrho_\theta)}},
\eea
%
where $\hat{\theta}$ expresses the unbiased estimator for $\theta$, and  $\nu$ is the number of times
the measurement is repeated.
Correspondingly, $\mathcal{F}(\varrho_\theta)$ is the so-called quantum Fisher information,  which is defined as~\cite{Helstrom,Holevo,Liu,Jin}
%
\begin{equation}
\label{eq17}
	\mathcal{F}(\varrho, A)=2\sum_{m,n}\frac{(p_m-p_n)^2}{(p_m+p_n)}|\langle m|A|n\rangle|^2,
\end{equation}
%
where $p_m$ and $|\phi_{m}\rangle$ represent the eigenvalues and eigenvectors of the
density matrix $\varrho$, respectively. 
Now, following the route provided in ref.~\cite{li2013}, the quantum Fisher information can  be written as
%
\begin{equation}
\label{eq18}
\mathcal{F}_{Q}=\sum_\mu \mathcal{F}(\varrho, A_\mu\otimes I+I\otimes B_\mu),
\end{equation}
%
where
$\{A_\mu\}$ and $\{B_\mu\}$
are arbitrary and natural complete sets of local orthonormal observables of the two subsystems with respect to $\varrho$.
The value of $\mathcal{F}_{Q}$ given by Eq.~(\ref{eq18}) is independent of the choice
of local orthonormal bases~\cite{li2013}, meaning that it is an inherent quantity of the composite system.
For a general two-spin system, the local orthonormal observables $\{A_\mu\}$ and $\{B_\mu\}$ can be defined as
%
\bea
\{A_\mu\}=\{B_\mu\}=\sqrt{2}\{I, S^x, S^y, S^z\},
\eea
%
and finally, for the reduced two-spin density matrix in Eq.~(\ref{eq4}) the analytical evaluation of the QFI
can be evaluated as
%
\be
\bl
\label{eq19}
&
\mathcal{F}_{Q}
=
\frac{
16
\Big(
\langle S_i^x S_{i+r}^x\rangle-\langle S_i^y S_{i+r}^y\rangle
\Big)^2
}
{1+4\langle S_i^z S_{i+r}^z\rangle}
\;
+
\\&
	\Bigg[
\frac{16}{
\Big(
1+
4\langle S_i^x S_{i+r}^x\rangle
\Big)
\Big(1+
4\langle S_i^y S_{i+r}^y \rangle
\Big)
-
4\langle S_i^z\rangle^2
}
\Bigg]\times
\\
&
\Bigg[
\!
\Big(
\!
3 \langle S_i^z\rangle^2
\!
+
\!
4\langle S_i^z S_{i+r}^z\rangle^2
\!\!
-
\!
2 \langle S_i^z S_{i+r}^z\rangle
\!
\Big)
\!
\Big(
\!
\langle S_i^x S_{i+r}^x\rangle
\!
+
\!
\langle S_i^y S_{i+r}^y\rangle
\!
\Big)
\\&\;\;
+
\frac{1}{2}
\Big(
\langle S_i^z\rangle^2
+
4\langle S_i^z S_{i+r}^z\rangle^2-
8 \langle S_i^z\rangle^2 \langle S_i^z S_{i+r}^z\rangle
\Big)
\\&\;\;
+
\Big(
1-
8 \langle S_i^z S_{i+r}^z\rangle
\Big)
 \Big(
 \langle S_i^x S_{i+r}^x\rangle^2+\langle S_i^y S_{i+r}^y\rangle^2
\Big)
\\&\;\;
+
4
\langle S_i^x S_{i+r}^x\rangle^3
+
4\langle S_i^y S_{i+r}^y\rangle^3
\Bigg] .
\el
\ee

%
\begin{figure}[b]
\centerline{
\includegraphics[width=\linewidth]{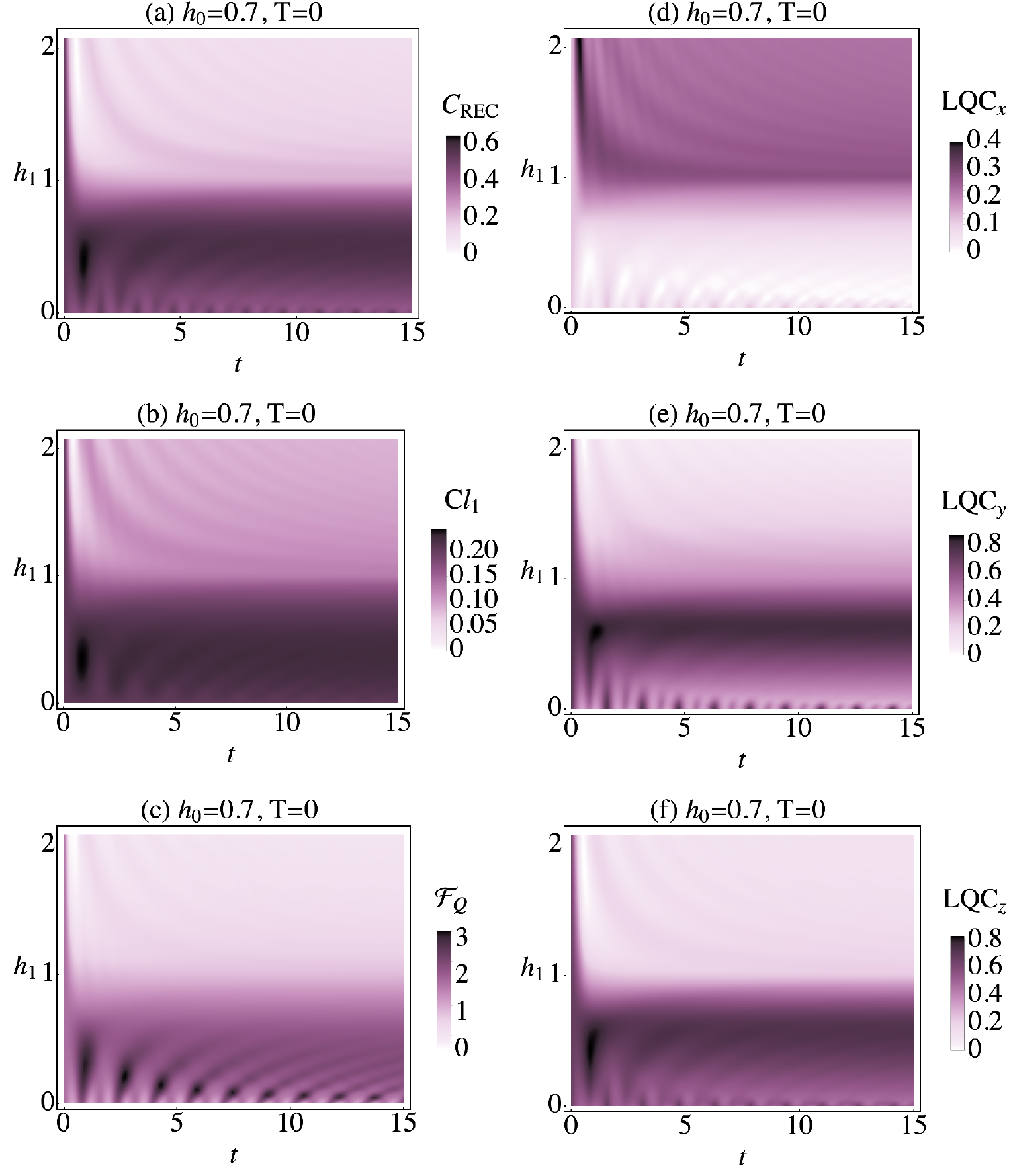}
}
\vspace{-0.3cm}
\caption{ (Color online)
Density plots of:
(a) the relative entropy of quantum coherence (${\rm C_{REC}}$),
(b) the $l_{1}$-norm of quantum coherence (Cl$_1$),
(c) the quantum Fisher information (QFI),
and (d-f) local quantum coherence (${\rm LQC_\alpha}$) with $\alpha=x,y,z$,
versus $t$ and $h_{1}$,  at zero temperature and for $h_{0}=0.7$ ($h_0<h_c$).
}
\label{fig1}
\end{figure}
%

%
\begin{figure}
\centerline{
\includegraphics[width=\linewidth]{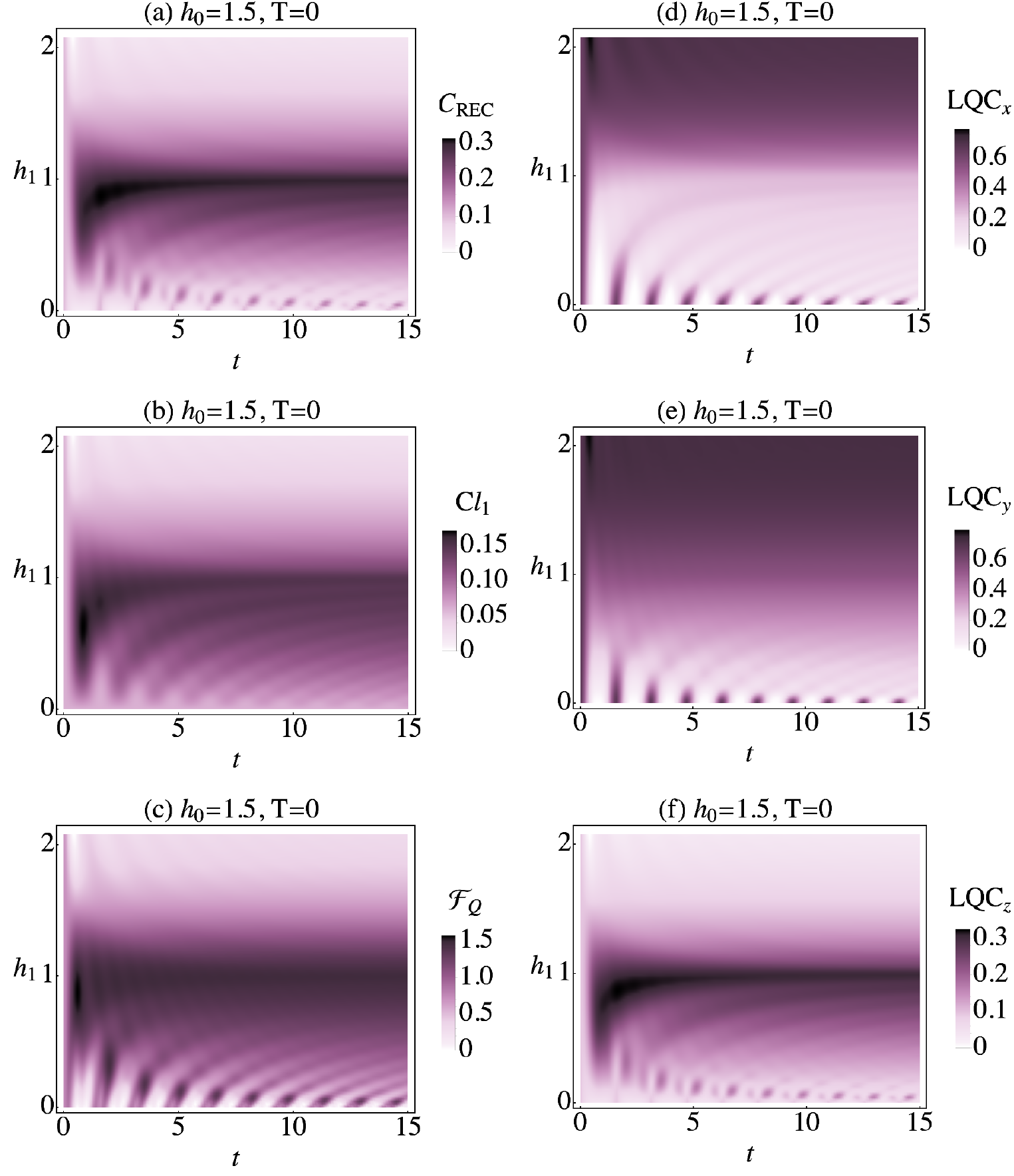}
}
\vspace{-0.3cm}
\caption{ (Color online)
Same density plots as Fig.~\ref{fig1} but for the case of  $h_{0}=1.5$ ($h_0>h_c$).
}
\label{fig2}
\end{figure}
%

%
\begin{figure}[b]
\centerline{
\includegraphics[width=\linewidth]{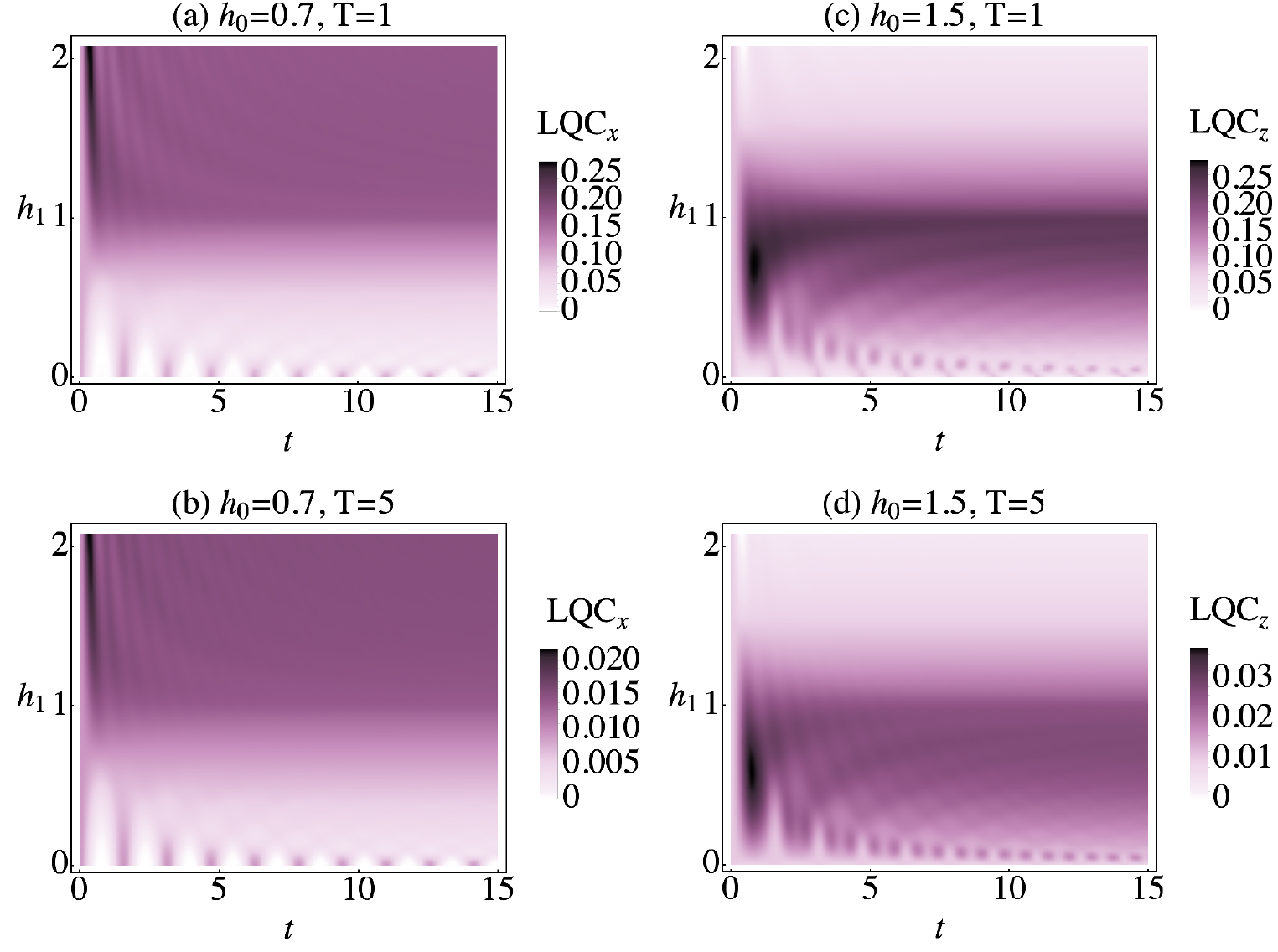}
}
\vspace{-0.4cm}
\caption{
(Color online)
The density plots of  the local quantum coherence versus $t$, and $h_{1}$,  for the  different temperatures of $T=1$ and $T= 5$.
 (a and b) show the ${\rm LQC_x}$  for $h_{0}=0.7$, and     (c and d) represent the ${\rm LQC_z}$ for $h_{0}=1.5$.
}
\label{fig3}
\end{figure}
%

%
\begin{figure*}
\includegraphics[width=0.99\linewidth]{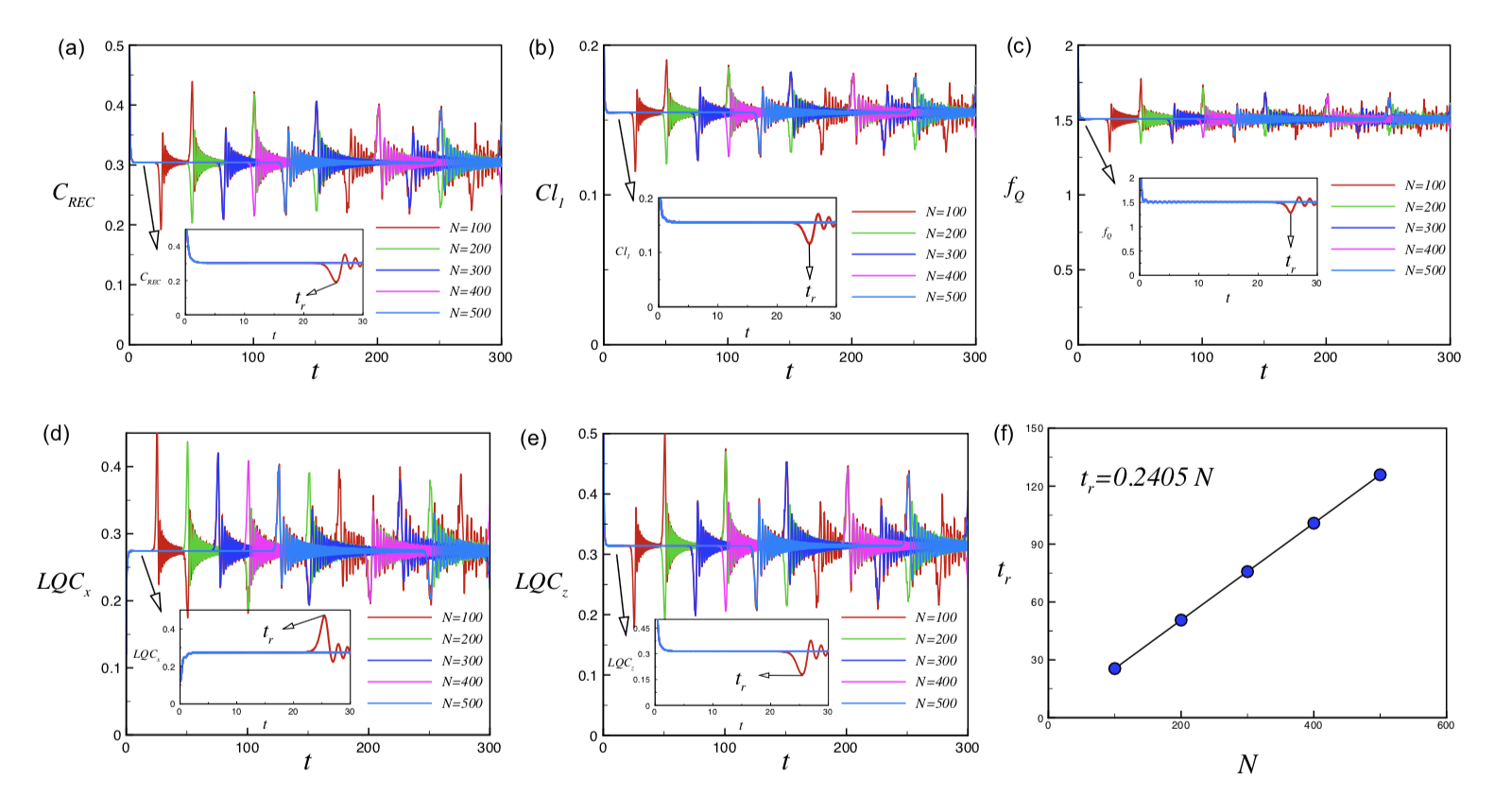}
\vspace{-0.4cm}
\caption{
(Color online)
The evolution of
(a)  the relative entropy of quantum coherence,
(b) the $l_{1}$-norm of quantum coherence
(c) the quantum Fisher information $\mathcal{F}_{Q}$,
and (d and e) local quantum coherence, for a quench to the critical
point $h_{1}=h_c$, for $h_{0}=0.7$ at zero temperature and for the  different system sizes.
(f) shows the linear  behaviour of  the first suppression-time (revival-time),  $t^{}_{r}(N)$,  versus the system size.}
\label{fig4}
\end{figure*}
%

%
\begin{figure*}
\begin{center}
\includegraphics[width=.97\linewidth]{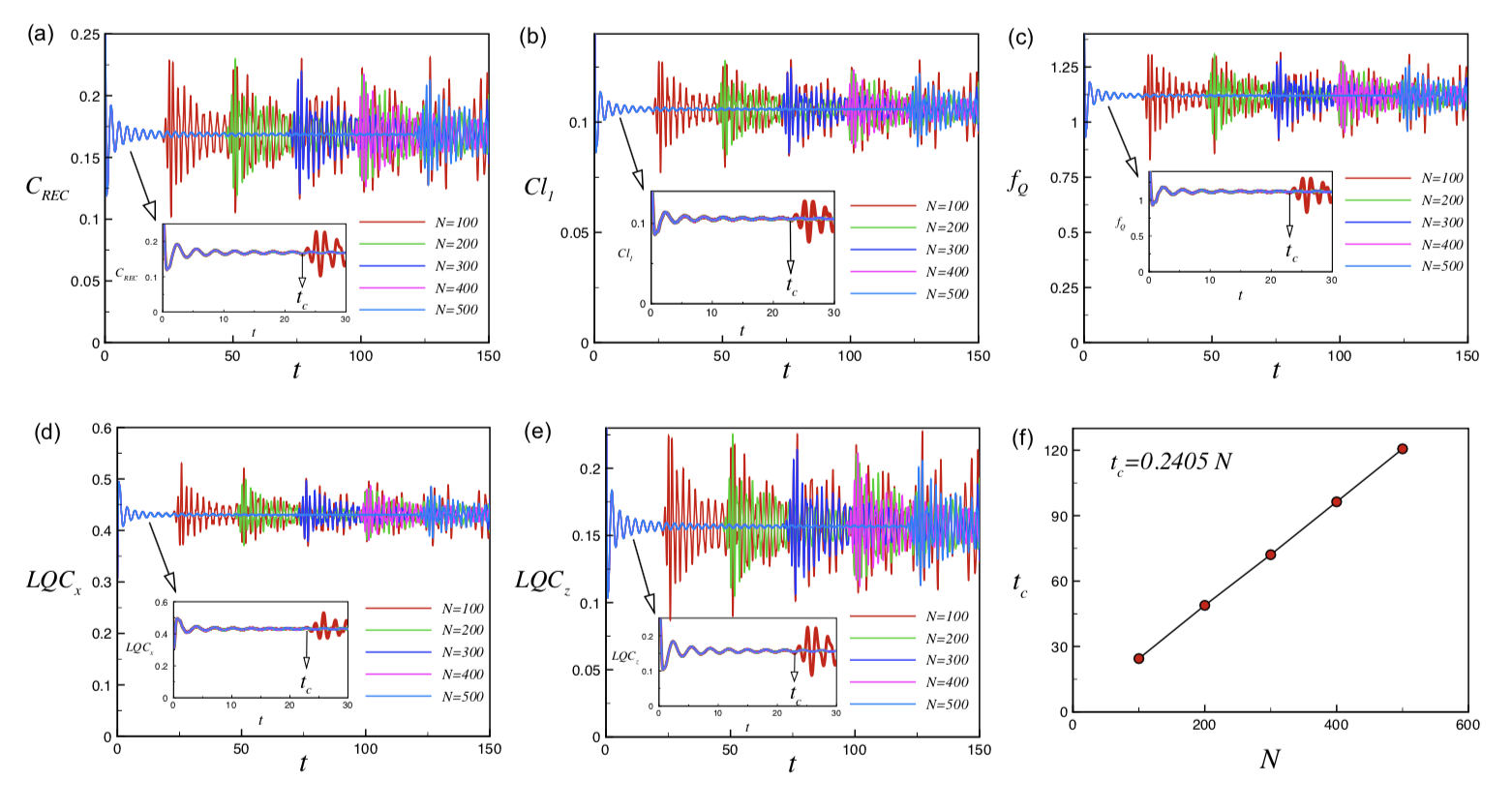}
\end{center}
\vspace{-0.3cm}
\caption{ (Color online)
(a-e) The same plots as Fig.~\ref{fig4}(a-e) but for the case that the system is at the critical point $h_{0}=h_c$, and quenched to $h_{1}=1.5$.
All plots shows again an oscillating behaviour after the time $t_c$, which scales linearly versus
 the system size shown in (f).
}
\label{fig5}
\end{figure*}
%

\section{Numerical Results and Discussions}
We now come to present our numerical results.
Although our formalism was for  a general case, for simplicity  we restrict our discussion  to the time dependent transverse magnetic field i.e., $J_{1}=J_{0}=1$.
Furthermore,  in the main text we only consider time dependent transverse filed Ising model (TFIM) by setting $\gamma=1$, 
and for more  general cases,  $\gamma\neq 1$,  one can look at the Appendix~\ref{AppB}.
It is well-known that the ground state of the TFIM is characterised by a quantum
phase transition that takes place at the critical point $h_{c}=J_0$~\cite{Pfeuty:1970aa,McCoy:1968aa}.
This phase transition is  a result of
the quantum fluctuations at zero temperature, which destroy the quantum correlations in the ground state.
It is determined via the order parameter, $\bra M^{x} \ket$, which differs from a finite value  for $h<h_{c}$ to  zero for $h\geq h_{c}$.
Moreover,  the ground state is  ferromagnetic aligned in $x$-direction for zero magnetic field
and it  has a  paramagnetic alignment along  the field for the limit of large magnetic field.
Both cases are minimally entangled since
the ground state   is a product of individual spin
states pointing in the $z$- ($x$-) direction    as $h \rightarrow \infty$ ($h \rightarrow 0$)~\cite{Pfeuty:1970aa,McCoy:1968aa}.
Furthermore,  by raising the temperature the entanglement shows a sudden decay near the critical point, although  at zero temperature and in a vicinity of critical point remains constant~\cite{Osborne:2002aa}.
\\

\subsection{Quench away from the critical point}
In Fig.~\ref{fig1}, we plot the intensity of the relative entropy of coherence (a),
the $l_{1}$-norm of coherence (b), the quantum Fisher information (c), and local quantum coherence components (d-f),
versus $t$ and $h_{1}$. The plots are   for $h_{0}=0.7$, and at zero temperature.
As one can see, for zero $h_{1}$, where the spins are completely aligned in the $x$-direction,
all quantities (expect C$l_{1}$) show an oscillatory behaviour in time.
By introducing  an external  magnetic field, $h_1$, the magnitude
of  quantities increases as  field increases until they reach their maximum value close to $h_{1}^{M}=h_{1}\approx0.5$.
In different circumstances, the maximum values of LQC$_{x}$ occurs at the equilibrium critical point $h_{1}^{M}=h_{1}=h_{c}$. 
As $h_{1}$ exceeds $h_{1}^{M}$, magnitude of all quantities decrease gradually by magnetic field.
Thus, when the system initially is prepared in ferromagnetic phase, $M^{x}(t, T=0)\neq0$, the maximum of two-spin
$S^{x}$ local coherence occurs at the equilibrium critical point and LQC$_{x}$ is the only quantity can capture truly
the critical point.
It should be mention that, when the system is prepared in its critical point, $h_{0}=h_c$, the maximum value that
quantities can reach is much greater than the previous case and appears at $h_{1}^{M}=h_{1}=h_{c}$.
\\

To further elaborate on the  behaviour of the zero temperature dynamics above the transition field, $h_0>h_c$,   Fig.~\ref{fig2} presents the intensity of the relative entropy of coherence (a),
the $l_{1}$-norm of coherence (b), the quantum Fisher information (c), and local quantum coherence components (d-f),
versus $t$ and $h_{1}$.
 We assume   $h_{0}=1.5$, where the system initially prepared at paramagnetic phase, $M^{z}(t,T=0) \neq 0$.
As seen, for  $h_{1}=0$, all quantities show an oscillatory behavior in time. Besides, when the external magnetic field is turned on ($h_{1}>0$),
the  magnitude of all quantities
except LQC$_{x}$ and LQC$_{y}$, enhances until they reach their maximum value at the equilibrium critical point $h_{1}^{M}=h_{1}=h_{c}$, then reduces by increasing the magnetic field.
From these findings one can conclude that  dynamical two-spin local $S^{z}$, quantum coherence (WYSI),
can positively pick out the  critical point while the REC, QFI, and C$l_{1}$ fail in this task.
To summarize: Depends on the initial state which the system is prepared, dynamics of the proper component of local quantum coherence can capture the critical point of the system. In other words, when the system is prepared in the initial state with $M^{\alpha}\neq0$, the dynamics of LQC$_{\alpha}$ reaches its maximum value at the critical point.
\\

On top of that, there would be a great interest to study the effect of temperature on the critical behavior of many body systems such as the spin systems~\cite{Sondhi:1997aa,Osborne:2002aa,Arnesen:2001aa,Gunlycke:2001aa}.
To show whether the WYSI is able to pinpoint the critical point at finite temperature, we  plot the LQC$_{x}$ and LQC$_{z}$
in Fig.~\ref{fig3} for different temperatures, namely $T=1,$~ and $T=5$.
Although the maximum value of LQC decreases as the temperature increases, the equilibrium phase transition point can still be signalled by the maximum of LQC$_{x}$ and LQC$_{z}$ at low temperature. This significant property can be easily applied to determine quantum critical points of the systems which today's technology makes it virtually impossible to achieve the necessary temperature that quantum fluctuations are dominated.
\\

\subsection{Quench from/to the critical point}
The time evolution of REC, C$l_{1}$, QFI, and LQC are  plotted for a quench to the critical
point $h_{1}=1$, for $h_{0}=0.7$ in Fig.~\ref{fig4}, for different system sizes.
As is clear,  in a very short time all quantities change rapidly from the equilibrium state to their average (constant) value that they oscillate around.
More than that, all quantities show
suppressions and revivals as deviations from the average value. In order to study the effect
of the system size on revival/suppression time, $t^{}_{r}$, we  also plot $t^{}_{r}(N)$ versus the system size in Fig. \ref{fig4}(f).
As seen, the $t^{}_{r}$ increases linearly by the system size, i.e.
\be
t_{r}(N)=\tau N,
\ee
where the scaling ratio is obtained as  $\tau=0.2405$.
A more detailed analysis shows that $t_{r}$ and $\tau$ are the same for all quenches and do not depend
on the  initial preparation phase of system. This is the promised universality of revival/suppression time, 
 which shows
that the size of the quench (different values of $h_{0}$) and the initial phase of system
are unimportant.

We also demonstrate  in Fig.~\ref{fig5}  the evolution of REC, C$l_{1}$, QFI, and LQC
for  $h_{1}=1.5$ and $h_{0}=h_c$, where the system prepared initially at the critical point.
Applying the external magnetic field causes a rapidly change in all quantities
from the equilibrium state to a constant value, before starting oscillations  at the time $t_c(N)$  [See the insets in Figs.~\ref{fig5}(a-e)].
In principle, $t_{c}(N)$ is an instances time under which all curves correspond to a  system larger than size $N$, clearly join together.
Examining the details in Fig.~\ref{fig5}(f), also shows  a  linear behavior of $t_{c}$ versus $N$,
\be
t_{c}(N)=\tau_c N,
\ee
and interestingly we find a similar scaling value as revival/suppression time, namely 
$\tau_c=\tau=0.2405$.
Our calculations show that $t_{c}$ and $\tau_c$ are the same for all quenches and do not depend
on the phase of system where it quenched to. This is the promised universality of $t_{c}$ which shows
that the size of the quench (different values of $h_{1}$) and the phase of system, where the system is quenched to,
are ineffectual.
\\

Finally, we  study the dynamics of REC, C$l_{1}$, WYSI and QFI for anisotropic case $\gamma\neq0$.
Our numerical analysis show that our previous findings are correct for anisotropic case (see Appendix~\ref{AppB}).
It is worthwhile to mention that, for the case $\gamma<0$, when the system is initialized at the phase with
$M^{y} ( t,T=0)\neq0$, the critical point of the system is signalled by the maximum of LQC$_{y}$ (see Appendix~\ref{AppB}).
Moreover, the numerical simulation shows that $t_{r}$, $t_{c}$ and their scaling ratios, $\tau$ and $\tau_c$ are
independent of the anisotropy parameter.

\section{Summary}
We have reported the dynamical behaviour of quantum coherence in the one dimensional time-dependent
transverse magnetic field XY-model.
For this purpose, we  investigate the dynamics of relative entropy of coherence, $l_{1}$-norm of coherence, Wigner-Yanase-skew information, and quantum Fisher information.
%
We  show that, the phase-transition point can be signalled by the maximum of %
Wigner-Yanase-skew information local components even at low temperature.
While %
relative entropy of coherence, $l_{1}$-norm of coherence and quantum Fisher information
lack such an indicator of criticality in the model.
In addition, we find that all of these quantities show suppressions and revivals by quenching the system  to the critical point.
Further, the first suppression (revival) time scales linearly with system size, and
free from  the quench size and the  initial phase of system, therefore our work highlights  the universality in out-of-equilibrium quantum many-body systems.
\\

The success of Wigner-Yanase-skew information dynamics to reveal the equilibrium phase transition may originate from its dependence on the square root of the elements of the density matrix. Therefore, It is a meaningful proposal to study the dynamics of similar quantifiers with a functionality of the square root of the elements of the density matrix.
Moreover, it will be interesting to extend the current investigation to more general time-dependent cases of the external magnetic field, such as exponential or periodic functions, and also it is worthwhile to extend the calculation to disorder case.

\section*{Acknowledgment}
We thank U. Mishra for helpful discussions.
A.A. acknowledges financial support from the National Research Foundation (NRF) funded by the Ministry of Science of Korea (Grants: No. 2016K1A4A01922028, No. 2017R1D1A1B03033465, and No. 2019R1H1A2039733).

\appendix
\section{Two point correlation functions of time dependent XY model}
\label{AppA}
%
%
%
%
%
Using Eqs.~(\ref{mag1}, and  \ref{mag2}) the expectation value of the magnetization along  the $z$-direction, specifically, is given by~\cite{Barouch1,Barouch2,Sadiek_2010,Huang2006,Utkarsh},
%
\begin{widetext}
\be
\bl
\label{eqApp3}
\langle M^{z}\rangle=
 \frac{1}{4N}\sum_{p=1}^{N/2}
 \frac{\tanh[\beta \Gamma(h_{0},J_{0})]}{\Gamma^2(h_{1},J_{1})\Gamma(h_{0},J_{0})}
\Bigg[
2 J_{1}(J_{0} h_{1} -J_{1} h_{0}) \delta_{p}^2 \sin^2[2t \Gamma(h_{1},J_{1})]
+4\Gamma^2(h_{1},J_{1})(J_{0}\cos\phi_{p}+h_{0})
\Bigg],
\el
\ee
%
where $\phi_{p}=2 \pi p/N$ , $\delta_{p}=2 \gamma \sin \phi_{p}$ and $\beta=1/K_B T$. Here  $K_B$ is Boltzmann constant and $T$ is the temperature.
One can simply use the Wick Theorem~\cite{Wick} to   obtain the nearest-neighbor spin correlation functions as follows
%
\be
\bl
\label{eqApp4}
\langle S^{x}_{l} S^{x}_{l+1} \rangle
=
\frac{1}{4}F_{l,l+1};
\;\;\;
\langle S^{y}_{l} S^{y}_{l+1} \rangle
=
\frac{1}{4}F_{l+1,l};
\;\;\;
\langle S^{z}_{l} S^{z}_{l+1} \rangle
=
\frac{1}{4}
\Big[
F_{l, l}\times F_{l+1,l+1}-Q_{l,l+1}\times G_{l,l+1}
-F_{l+1,l}\times F_{l,l+1}
\Big],
\el
\ee
%
in which by defining
%
$\Gamma[h(t),J(t)]=\Big[
J(t)\cos \phi_{p} + h(t)]^{2}+\gamma^2 J^2(t) \sin^2\phi_{p}
\Big]^{\frac{1}{2}}$, we can write
%
\be
\label{eqApp5}
\bl
Q_{l, m}&=\frac{1}{N} \sum_{p=1}^{N/2}
\Bigg[
2\cos[(m-l)\phi_{p}]
+\frac{i(J_{1}h_{0}- J_{0}h_{1}) \delta_{p} \sin[(m-l)\phi_{p}]\sin[4t\Gamma(h_{1},J_{1})]\tanh[\beta \Gamma(h_{0},J_{0})]}{\Gamma(h_{1},J_{1})\Gamma(h_{0},J_{0})}
\Bigg]
,\quad
\\
 G_{l, m}&=\frac{1}{N} \sum_{p=1}^{N/2}
 \Bigg[
-2\cos[(m-l)\phi_{p}]
+\frac{i(J_{1}h_{0}- J_{0}h_{1}) \delta_{p} \sin[(m-l)\phi_{p}]\sin[4t\Gamma(h_{1},J_{1})]\tanh[\beta \Gamma(h_{0},J_{0})]}{\Gamma(h_{1},J_{1})\Gamma(h_{0},J_{0})}
\Bigg]
,\quad
\el
\ee
%
%
\be
\bl
F_{l, m}=
&
\frac{1}{N} \sum_{p=1}^{N/2} \frac{\tanh[\beta \Gamma(h_{0},J_{0})]}{\Gamma^2(h_{1},J_{1})\Gamma(h_{0},J_{0})}
\Bigg[
\cos[(m-l)\phi_{p}]
\times
\Big[
J_{1} [J_{0} h_{1} - J_{1} h_{0}] \delta^2_{p} \sin^2[2t\Gamma(h_{1},J_{1})] + 2\Gamma^2(h_{1},J_{1})(J_{0}\cos\phi_{p}+h_{0})
\Big]
\\
&+ \delta_{p} \sin[(m-l)\phi_{p}]
\Big[
J_{0}\Gamma^2(h_{1},J_{1}) +2 (J_{1} h_{0} - J_{0} h_{1}) (J_{1}\cos\phi_{p}+h_{1}) \sin^2[2t\Gamma(h_{1},J_{1})]
\Big]
\Bigg].
\el
\ee
%
\end{widetext}

\section{WYSI for anisotropic case $\gamma\neq0$}{\label{AppB}}

%
\begin{figure}
\vspace{0.3cm}
\centerline{
\includegraphics[width=\linewidth]{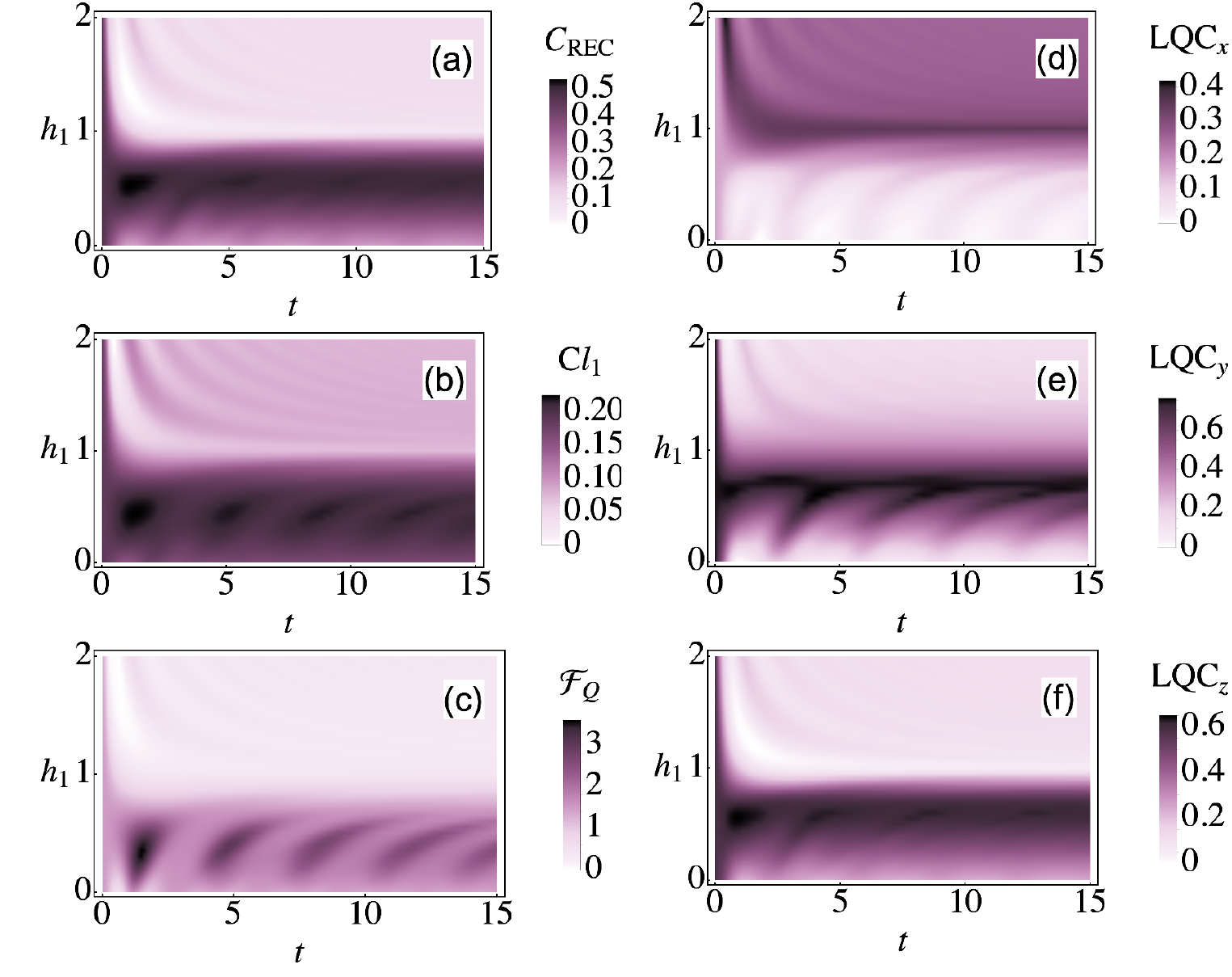}
}
\vspace{-0.3cm}
\caption{ (Color online)
Same density plots as Fig.~\ref{fig1} but for the case of $\gamma=0.5$,   at zero temperature and for $h_{0}=0.7$ ($h_0<h_c$).
}
\label{Fig1AppB}
\end{figure}
%
%
\begin{figure}
\vspace{0.3cm}
\centerline{
\includegraphics[width=\linewidth]{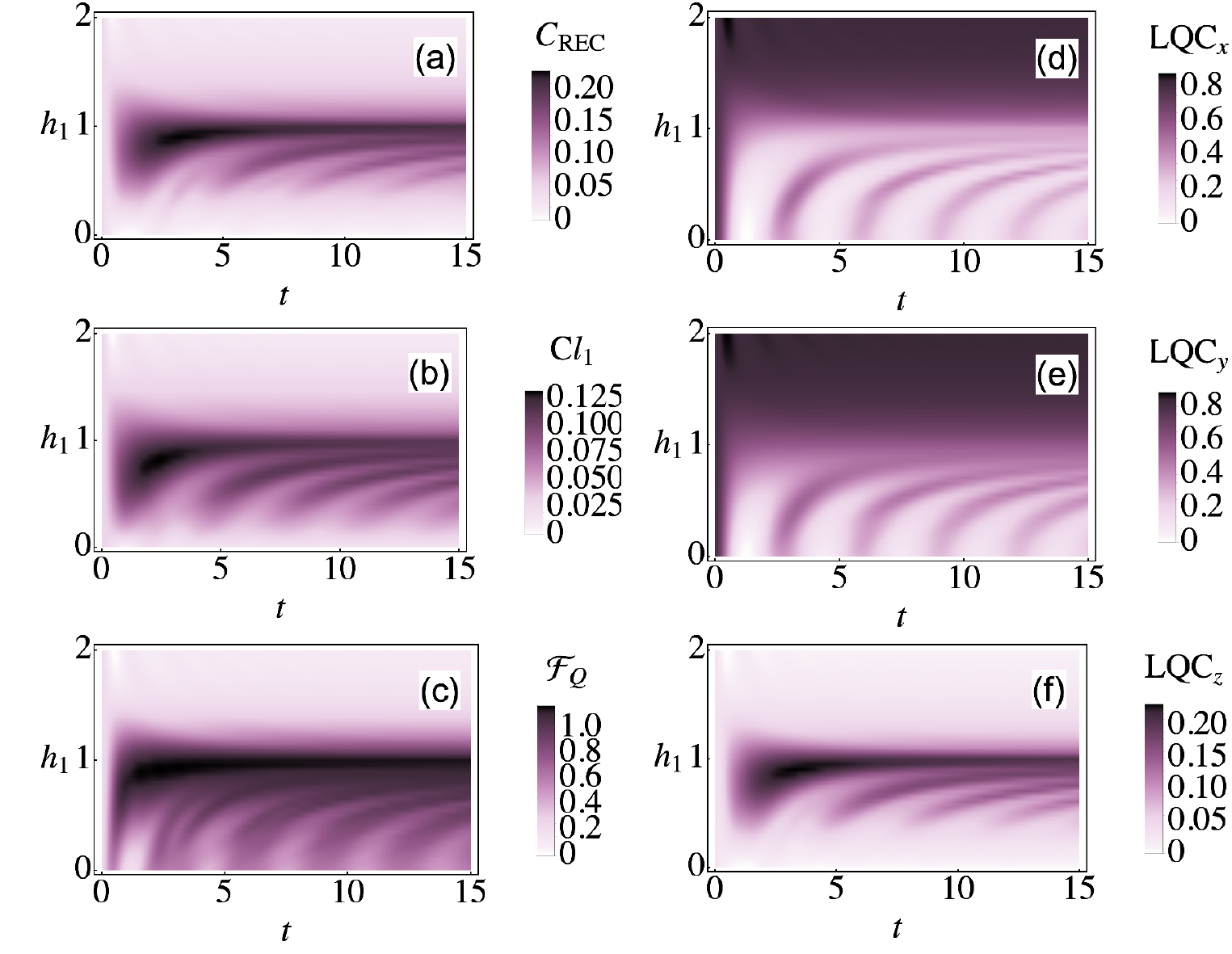}
}
\vspace{-0.3cm}
\caption{ (Color online)
Same density plots as Fig.~\ref{fig1} but for the case of  $\gamma=0.5$ at zero temperature and  $h_{0}=1.5$ ($h_0>h_c$).
}
\label{Fig2AppB}
\end{figure}
%
%
\begin{figure}
\vspace{0.3cm}
\centerline{
\includegraphics[width=\linewidth]{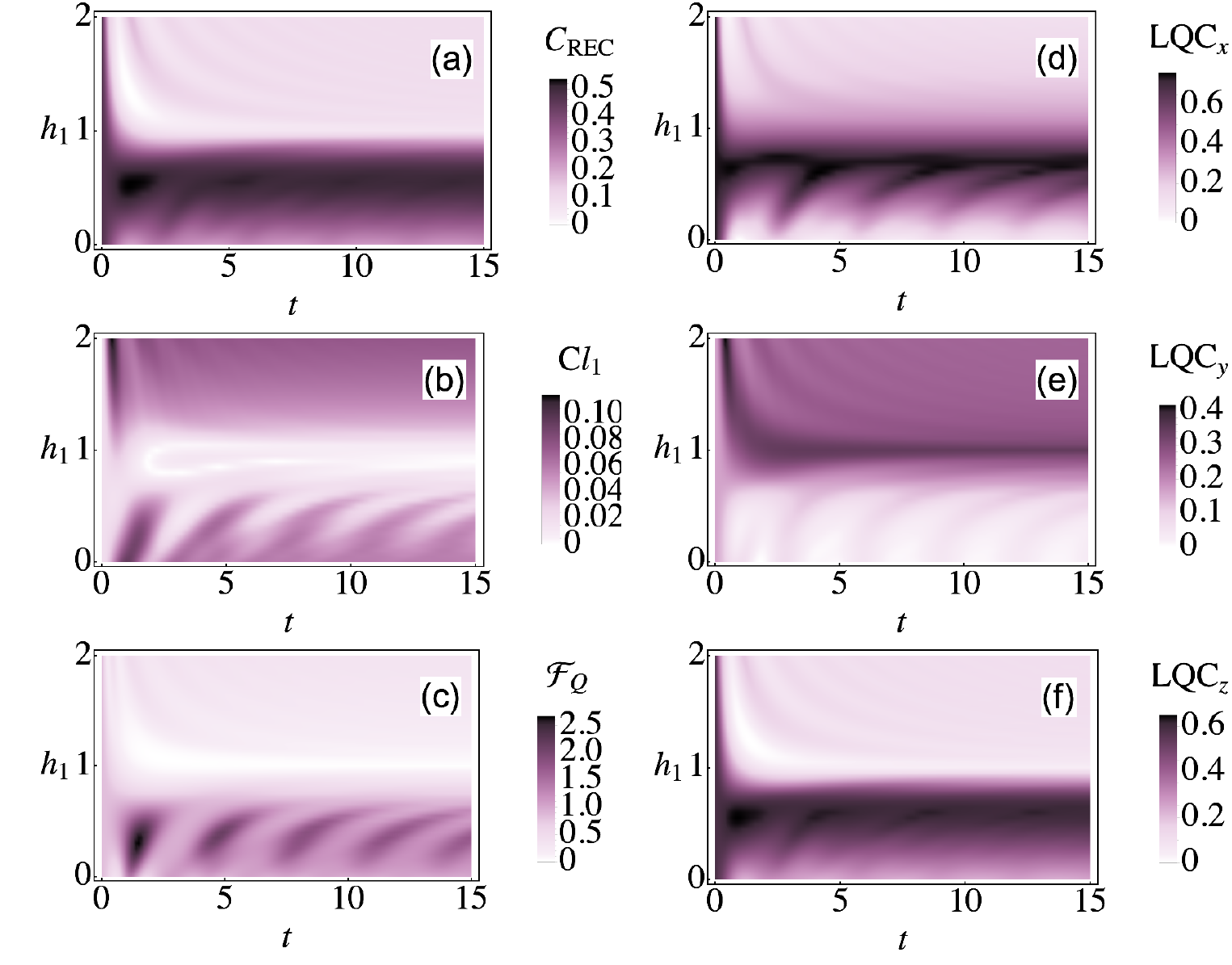}
}
\vspace{-0.3cm}
\caption{ (Color online)
Same density plots as Fig.~\ref{fig1} but for the case of  $\gamma=-0.5$ at zero temperature  and  $h_{0}=0.7$ ($h_0<h_c$) with magnetisation along the $y$-direction.
}
\label{Fig3AppB}
\end{figure}
%
In this appendix, we study the dynamics of the quantities for anisotropic case $\gamma\neq0$ (Eq.~\ref{eq1}). For this purpose  in the Fig.~\ref{Fig1AppB},  we  first look at the  anisotropic case $\gamma=0.5$. We show
the density plot of dynamical behaviour of the relative entropy of coherence
(a), the $l_{1}$-norm of coherence (b), the quantum Fisher information (c), and local quantum coherence components (d-f),
versus time and $h_{1}$, for $J_{0}=J_{1}=1$, and $h_{0}=0.7$.
As we expect, when the initial state prepared in ferromagnetic case $M^{x}\neq 0$, the LQC$_{x}$ shows maximum at the critical point of the system $h_{c}=1$.
Moreover, we show that when the system initialized at paramagnetic phase $M^{z}\neq 0$, i.e., $h_{0}>1$,  the LQC$_{z}$ fulfils the expectation and reaches its maximum at the critical point.
This is clearly represented in the results of Fig.~\ref{Fig2AppB}.
Finally, for the case that, the system initially prepared at ferromagnetic phase $\gamma<0$, in which $M^{y} \neq 0$, the maximum of LQC$_{y}$ happens
 at the critical point of the system (see Fig.~\ref{Fig3AppB}).
Briefly, one can conclude that when the system is prepared in the initial state with $M^{\alpha} \neq 0$, the dynamics of LQC$_{\alpha}$ reaches its maximum value at the critical
point.

%


\end{document}